\documentstyle[aps,12pt,draft]{revtex}

\begin{document}
\draft
\title{Dc-Driven Diatomic Frenkel-Kontorova Model}
\author{Aiguo Xu and Jongbae Hong}
\address{Department of Physics and Center for Strongly Correlated \\
Materials Research, Seoul National University, \\
Seoul 151-742, Korea}
\maketitle

\begin{abstract}
We investigate the resonance steps, spatiotemporal dynamics, and dynamical
phase diagrams of the dc-driven diatomic Frenkel-Kontorova model. The
complete resonance velocity spectrum is given. The diatomic effects result
in each resonant state being characterized by two integer pairs, $(k_1,k_2)$
and $(k_1,k_2^{\prime })$. In the high-velocity region the linear response
of the average velocity , $v$, of the chain to the driving force $F$ is
often punctuated by subharmonic resonances $(k_1>k_2)$. There are two kinds
of nonlinear response regions in the high-velocity region. A new physical
interpretation of the mean-field treatment is presented.
\end{abstract}


\newpage

The static and the transport properties of Frenkel-Kontorova(FK) \cite{1}
systems have been subjects of intensive studies\cite{2,3,4}. In these
studies, one of the most important parameters is the winding number $\omega $%
\cite{5}. Starting from uniform sliding states, Strunz and Elmer\cite{6}
presented the main, superharmonic resonant conditions while Zheng et al\cite
{7} gave the complete resonance condition. The main studies in Refs.\cite{6}
and \cite{7} are limited to the region of low velocity and to the case where
the chain is composed of the same atoms.

In this paper, we study the dc-driven FK system where the chain is composed
alternatively of two different kinds of atoms. The response of the average
velocity of the system to the driving force, the resonant steps, and the
spatiotemporal dynamics are studied analytically and numerically. Our
studies shed some different light on the dynamical behavior of the driven FK
systems because (i) diatomic effects which result in new resonance modes are
shown; (ii) regions far away from those addressed in previous investigations
are treated; and (iii) a new physical interpretation of the mean-field
treatment is presented.

We consider a uniform, quasi-static force field $F(t)$ acting on a FK system
in the presence of a damping proportional to the velocity and where the
atoms can exchange energy with a thermal bath. When the FK chain is composed
of more than one kind of atoms, the motion equation of the system may be
expressed as

\begin{eqnarray}
m_j\stackrel{..}{x_j} &=&-m_j\eta \stackrel{.}{x}_j+(x_{j-1}-2x_j+x_{j+1}) 
\nonumber \\
&&-\frac K{2\pi }\sin (2\pi x_j)+F\text{ .}  \label{eq1}
\end{eqnarray}
In this paper, we choose $m_1=m_3=\cdots =1$, $m_2=m_4=\cdots =m$ where $%
m\geq 1$. The ground states of this diatomic FK model have been studied in
one of our previous papers\cite{8}.The diatomic effect results in a second
critical point.

When $K=0$, there is no nonlinear interaction in the system, and the system
mobility shows a linear response $B=1/\eta $. When $K>0$, the interactions
between the atoms and the external potential are nonlinear. The commensurate
ground states are always pinned. For an incommensurate ground state with the
winding number $\omega $, there exists a critical value $K_C(\omega )$ at
which the Aubry transition occurs\cite{9}. When $K<K_C(\omega )$,  the
system is unpinned; when $K>K_C(\omega )$, the system is pinned. For the
latter case, when $F$ is less than the depinning force $F_d$, the system is
in the locked state. When $F>K/(2\pi )$, the system should behave
approximately as a homogeneous one, and $B(F)$ reaches the final value $%
B^f=1/\eta $ (linear response). For an intermediate force, the mobility
depends on the driving strength (nonlinear response).

The average sliding velocity of the chain reads

\begin{equation}
v=\lim_{T\rightarrow \infty }\frac 1T\int_0^T\frac 1N\sum_{j=1}^N\dot{x}_jdt%
\text{.}  \label{eq2}
\end{equation}
For an incommensurate structure with winding number $\omega $, when $%
K<K_C(\omega )$ at which the transition by breaking of analyticity occurs 
\cite{9}, $\frac 1N\sum_{j=1}^N\sin (2\pi x_j)=0$, so in the case of no
dissipation, the chain may slide without any applied force. If the particles
are charged, this state corresponds to the superconductivity state.

For a numerical solution to Eq. (\ref{eq1}), we use the standard
fourth-order Runge-Kutta method. Periodic boundary conditions, $x_{N+1}=x_1+M
$, $x_0=x_N-M$, $\stackrel{.}{x}_0=\stackrel{.}{x}_N$, $\stackrel{.}{x}%
_{N+1}=\stackrel{.}{x}_1$, are imposed, where $M$, $N$ are positive integers
and $N$ is an even number. In the simulation, we start with the ground state
and $F=0.$ We increase the dc force $F$ by small steps $\Delta F$ up to a
value $F_{\max }$. Then, we decrease it to zero with the same steps. At each
step, we wait long enough to allow the system to reach a stationary state.
Then, for the discrete times $t_i=i\Delta t$, we measure system
characteristics such as the average velocity $<v>$.

The symmetries of the equation of motion lead to the existence of
nonstationary solutions called uniform sliding states. They are
characterized by the fact that every odd-numbered (or even-numbered) atom
performs the same motion, but is shifted in time. That is,  $%
x_{2n+1}(t)=x_{2n-1}(t+T_1)$ (or $x_{2(n+1)}(t)=x_{2n}(t+T_2)$) for $%
n=1,2,\cdots $. It is clear that $T_1=T_2$, $v_1=v_2=v$, where $v_1$ and $v_2
$ are the average velocities of the odd-numbered and the even-numbered
atoms, respectively. Thus, we need only two dynamical hull functions, $f_1$
and $f_2$, to describe the motions of all atoms:.

\begin{eqnarray}
x_{2n-1}(t) &=&\psi _0+\psi +(2n-1)\omega +vt+  \nonumber \\
&&f_1[\psi +(2n-1)\omega +vt],  \label{eq3}
\end{eqnarray}
\begin{eqnarray}
x_{2n}(t) &=&\psi +2n\omega +vt+  \nonumber \\
&&f_2(\psi +2n\omega +vt),  \label{eq4}
\end{eqnarray}
where $\psi _0+\psi +(2n-1)\omega +vt$ and $\psi +2n\omega +vt$ are the
positions of the $(2n-1)$th and $(2n)$th atoms for the case of $K=0$. The
dynamical hull functions $f_1$ and $f_2$ describe the modulations of the
external potential to the atomic positions, $\psi $ is an arbitrary phase,
and $\psi _0$ is a constant phase. Because of the discrete translation
symmetry of Eq. (\ref{eq1}), the hull functions are periodic, and their
periods may be any positive integer $n_1$; i.e.

\begin{equation}
f_1(\varphi _{2n-1}+n_1)=f_1(\varphi _{2n-1}),  \label{eq5}
\end{equation}

\begin{equation}
f_2(\varphi _{2n}+n_1)=f_2(\varphi _{2n}).  \label{eq6}
\end{equation}
It should be noted that $f_1(\varphi _{2n-1}+n_1-1)\neq f_1(\varphi _{2n-1})$
and $f_2(\varphi _{2n}+n_1-1)\neq f_2(\varphi _{2n})$ when we say the period
is $n_1$. That is to say, uniform sliding states have infinite kinds of
modes. This fact is of key importance in obtaining the complete resonance
condition.

In the frame of the mass center which moves with the average sliding
velocity $v$, the external potential leads to a time-periodic force acting
on the atoms. The frequency of this force, the so-called ``washboard
frequency,'' is given by the velocity of the mass center divided by the
period of the potential; i.e., $\Omega _w=2\pi v$ . A resonance may occur if
the washboard frequency $\Omega _w$ and the eigenfrequencies $\Omega _{p\pm }
$ of the phonon satisfy the condition $k_1\Omega _{p+}=k_2\Omega _w$ or $%
k_1\Omega _{p-}=k_2\Omega _w$, where $k_1$ and $k_2$ are positive coprime
integers. To see this and to find the eigenfrequencis $\Omega _{p\pm }$ of
the phonon, we (i) use the linear approximations $\sin \{2\pi [\varphi
_{2n-1}+f_1(\varphi _{2n-1})]\}\approx \sin (2\pi \varphi _{2n-1})+2\pi \cos
(2\pi \varphi _{2n-1})f_1(\varphi _{2n-1})$ and $\sin \{2\pi [\varphi
_{2n}+f_2(\varphi _{2n})]\}\approx \sin (2\pi \varphi _{2n})+2\pi \cos (2\pi
\varphi _{2n})f_2(\varphi _{2n})$, (ii) introduce the mean-field treatment,
i.e., the term $\cos (2\pi \varphi _j)$ is replaced by an averaged quantity

\begin{equation}
\beta =\lim_{T\rightarrow \infty }\frac 1T\int_0^T\frac 1N\sum_{j=1}^N\cos
(2\pi \varphi _j)dt,  \label{eq7}
\end{equation}
(iii) consider a small vibration $\varepsilon _j(t)$ of the atoms around
their positions, and (iv) use the Fourier expansion

\begin{equation}
f_1(\varphi )=\sum_{n_2}C_{n_2}\exp (i\frac{2\pi n_2}{n_1}\varphi )\text{, }
\label{eq8}
\end{equation}
\begin{equation}
f_2(\varphi )=\sum_{n_2}D_{n_2}\exp (i\frac{2\pi n_2}{n_1}\varphi )\text{. }
\label{eq9}
\end{equation}
Finally, we obtain the resonance velocity

\begin{equation}
v_{\pm }=\frac{n_1}{2\pi n_2}{\LARGE \{}\frac 1{2m}[(2+K\beta )(1+m)\pm
\gamma ]{\LARGE \}}^{1/2},  \label{eq10}
\end{equation}
where $\{\frac 1{2m}[(2+K\beta )(1+m)\pm \gamma ]\}^{1/2}=\Omega _{p\pm }$, $%
\gamma =[(2+K\beta )^2(m-1)^2+16m\cos ^2(\frac{2\pi n_2}{n_1}\omega )]^{1/2}$%
, and $2\pi v_{\pm }=\Omega _w$. It is clear that $k_1/k_2=n_1/n_2$. This is
a complete resonance velocity spectrum which relates only to the winding
number, i.e., the commensurability, implies that the resonance is not a
finite-size effect, but a discreteness effect. From Eq. (\ref{eq10}), it is
clear that the resonance behavior of the present model is very different
from that of the  standard driven FK model. For a given integer pair $%
(k_1,k_2)$, the present model has two possible resonance states. For a given
resonance state, the value of the average velocity $v$ is fixed, so if the
value of $k_1$ is fixed, then $k_2$ should have two possible values which
correspond to the two cases where the  ``$+$'' and ``$-$'' signs are taken.
So one resonance state should be characterized by two integer pairs, $%
(k_1,k_2)$ and $(k_1,k_2^{\prime })$. That is to say, there are two kinds of
resonance modes in one state.

When $m=1$, we can reduce Eq. (\ref{eq10}) to a simpler form: 
\begin{eqnarray}
v_{+} &=&\frac{n_1}{2\pi n_2}\sqrt{K\beta +4\cos ^2(\frac{\pi n_2}{n_1}%
\omega )}\text{,}  \nonumber \\
v_{-} &=&\frac{n_1}{2\pi n_2}\sqrt{K\beta +4\sin ^2(\frac{\pi n_2}{n_1}%
\omega )}\text{,}  \label{eq11}
\end{eqnarray}
where $v_{-}$ is the real resonance velocity while $v_{+}$ is a fake one.
From Eqs. (\ref{eq10}) and (\ref{eq11}), we can clearly find that the
external potential not only leads to a periodic driving force but also leads
to a modulation to the eigenfrequencies. In the high-velocity region, the
standard FK model shows a behavior similar to that of the diatomic FK model.
It is clear that $E_p^{\prime }=-\beta /(4\pi ^2)=$ $\lim_{T\rightarrow
\infty }\frac 1T\int_0^TE^{\prime }(p)dt$ is the average external potential
energy experienced by an atom. In the following, we will see that the
dependence of $E_p^{\prime }$ on $F$ helps us to understand better the
resonance steps .

Figure 1 shows the response of the average velocity $v$ of the system to the
driving force $F$, where $\omega =3/8$, $\eta =0.1$, $K=1$, $0<F<0.1$, $%
\Delta F=0.001$, and the integral step $\Delta t=0.01$. The line with solid
squares and the dotted squares without a line are for the case of $m=2$. The
line with solid circles and the dotted circles without a line are for the
standard FK model which is plotted for comparision. It is clear that for the
diatomic FK model, each resonance state is characterized by two pairs of
integers, ($k_1$,$k_2$) and ($k_1$,$k_2^{\prime }$), which are for small and
large atoms, respectively. The main property of a resonance step is
characterized by one kind of resonance state, but some other resonance
states may occur at the edges of the step. The ratio $k_1/k_2$ is always
smaller than $k_1/k_2^{\prime }$, which implies that the frequency of the
small atom is larger than that of the large one. Figure. 2 shows an example
of the corresponding dynamical behavior in a resonance state for the case of 
$m=2$, where $F=0.015$. The thin line is for small stoms and the thick line
is for large ones. The high peaks are due to hopping from one well to
another well, and the low peaks are due to neighboring-particle hopping. In
this case, the time between two neighboring high peaks for the thin line (or
the thick line) equals the period of the washboard, and there are four small
oscillations for the thin line and three small oscillations for the thick
line, so the resonance state is characterized by $(1,4)$, $(1,3)$.

Numerical results show that in the high-velocity region, the linear response
of $v$ to $F$ is often punctuated by nonlinear response regions. Figure.3
shows an example for the response of $v$ to $F$ in the high velocity region,
where only the process of increasing $F$ is shown: $m=2$, $\omega =1/8$, $%
\eta =0.1$, $K=1$, $\Delta F=0.002$, and the integral step $\Delta t=0.1$.
It is clear that the linear response of $v$ to $F$ is punctuated by
resonance steps which result from sub-harmonic resonances ( $k_1>k_2$). The
distances between two successive nonlinear response regions are nearly
equal. A nonlinear response region is usually composed of one large step and
several smaller ones. The main resonance properties are characterized by the
large step. There are two kinds of nonlinear response regions in the
high-velocity region. The two kinds of nonlinear response regions occur
alternatively.

To understand better the resonant steps in the high-velocity region, in
Fig.4 we show the average external potential energy $E_p^{\prime }$
experienced by an atom. The values of the parameters are same as those used
for Fig. 3. From Figs. 4(a) and 4(b), we can easily find that $E_p^{\prime }$
usually oscillates and has different properties for the two kinds of
nonlinear responses, which correspond to two kinds of dynamical
configurations of the moving chain. In one kind of the nonlinear response,
the time-averaged value of $E_p^{\prime }$ is higher than that in its
vicinity. In the other kind of nonlinear response, the time-averaged value
of $E_p^{\prime }$ is lower than that in its vicinity. Other interesting
phenomena are the facts that the average kinetic energy $E_k$, the average
interparticle potential $E_p$, and $E_p^{\prime }$ are all constants in the
cases of $F=1.5$, $3.0$, $4.5$, $6.0$, $\cdots $, which implies that the
moving chain does not feel the existence of the external potential in these
cases. The results given in this letter also can be used to describe an
dc-biased one-dimensional Josephson-junction array.

\acknowledgments 
This work was supported by Korea Research Foundation Grant
(KRF-2000-DP0106). The authors also wish to acknowledge the support by the
BK21 project of the Ministry of Education, Korea.

{\bf Fig. 1.} Response of the average velocity $v$ of the system to the
driving force $F$ for increasing $F$, where $\omega =3/8$, $\eta =0.1$, $K=1$%
, $0<F<0.1$, $\Delta F=0.001$, and the integral step $\Delta t=0.01$.

{\bf Fig. 2.} Dynamical behavior in a resonance state, where $m=2$, $F=0.015$%
. The thin line is for small stoms, and the thick line is for large ones.
The high peaks are due to hopping from one well to another well, and the low
peaks are due to neighboring-particle hopping. In this case, the time
between two neighboring high peaks for the thin line (or the thick line)
equals the period of the washboard. There are four small oscillations for
the thin line and three small oscillations for the thick line, so the
resonance state is characterized by $(1,4)$, $(1,3)$.

{\bf Fig. 3.} Response of the average velocity $v$ of the system to the
driving force in the high velocity region, where only the process of
increasing $F$ is shown, $m=2$, $\omega =1/8$, $\eta =0.1$, $K=1$, $\Delta
F=0.002$, and the integral step $\Delta t=0.1$.

{\bf Fig. 4.} Average external potential energy $E_p^{\prime }$ experienced
by an atom. The values denoted in the figure are for the external force. The
values of the parameters aresame as those used for Fig. 3.


\begin{references}
\bibitem{1}  T. Kontorova and Y. I. Frenkel, Zh. Eksp. Teor. Fiz. {\bf 8},
89 (1938); {\bf 8}, 1340 (1938); {\bf 8}, 1349 (1938).

\bibitem{2}  L. M. Flor\'{i} a and J. J. Mazo, Adv. Phys. {\bf 45}, 505
(1996), and references therein.

\bibitem{3}  O. M. Braun and Yu. S. Kivshar, Phys. Rep. {\bf 306}, 1 (1998),
and references therein.

\bibitem{4}  Aiguo Xu, Guangrui Wang, Shigang Chen, and Zhanru Yang, Prog.
Phys. (Nanjing, CHINA){\bf 19}, 109 (1999), and references therein.

\bibitem{5}  Aiguo Xu, Guangrui Wang, Shigang Chen, and Bambi Hu, Phys. Rev.
B {\bf 57}, 2771 (1998); Phys. Lett. A {\bf 233}, 99 (1997). For the
standard FK model, the winding number $\omega =M/N$, where $M$ and $N$ are
integers, $N$ is the period of the chain, $M$ is the number of the period of
the external potential between the first and the $(N+1)$th atoms. For the
diatomic FK model, if $F_{n+1}$ is an odd number, then $\omega
=(2F_n)/(2F_{n+1})$.

\bibitem{6}  T. Strunz and F. J. Elmer, Phys. Rev. E {\bf 58}, 1601 (1998); 
{\bf 58}, 1612 (1998).

\bibitem{7}  Z. Zheng, B. Hu, and G. Hu, Phys. Rev. B {\bf 58}, 5453 (1998).

\bibitem{8}  Aiguo Xu, Guangrui Wang, Shigang Chen, and Bambi Hu, Phys. Rev.
B {\bf 58}, 721 (1998).

\bibitem{9}  S. Aubry, Physica D {\bf 7}, 240 (1983); S. Aubry and P. Y. Le
Daeron, Physica D {\bf 8}, 381(1983); L. de Seze and S. Aubry, J. Phys. C 
{\bf 17}, 389 (1984).

\newpage 
\end{references}
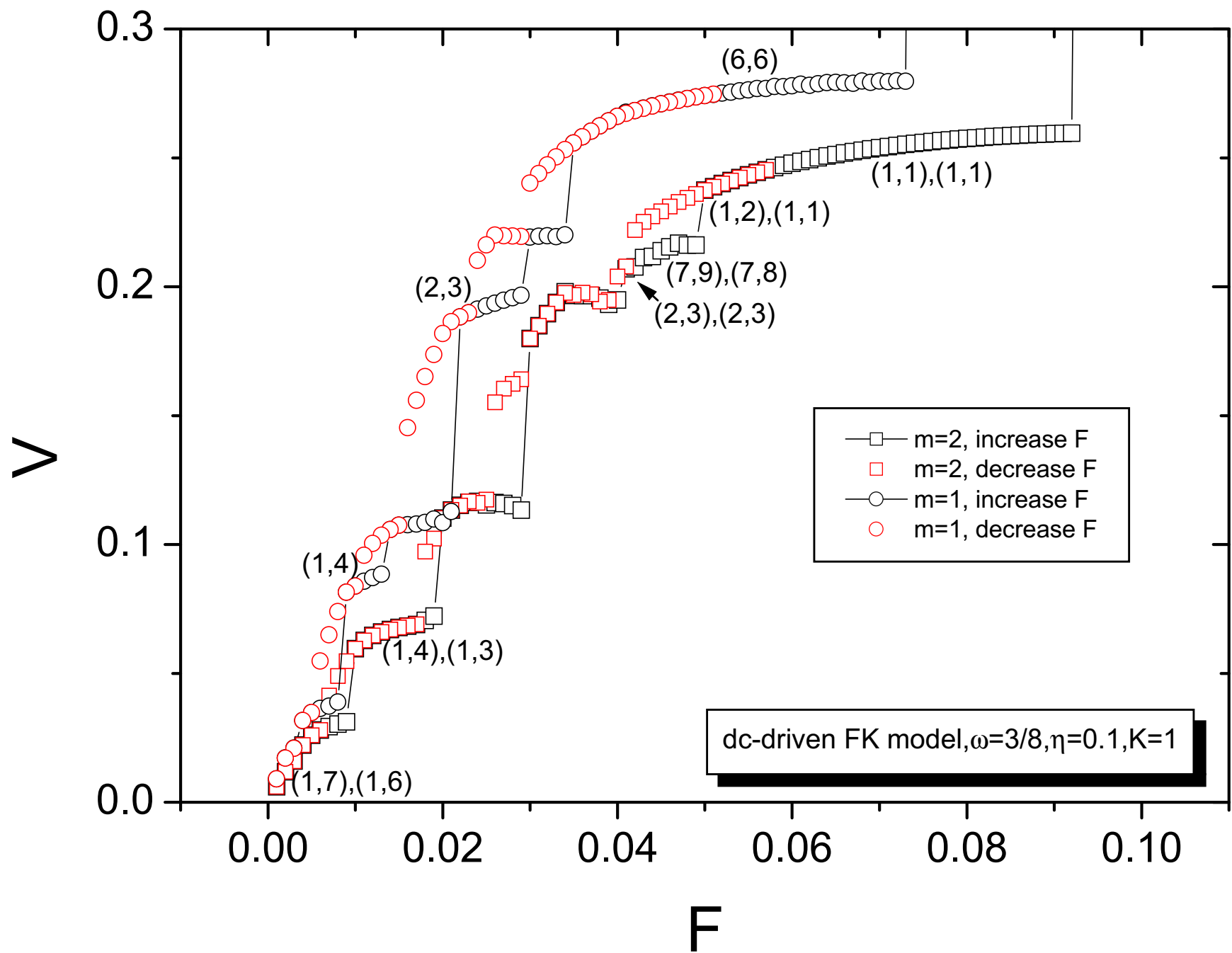
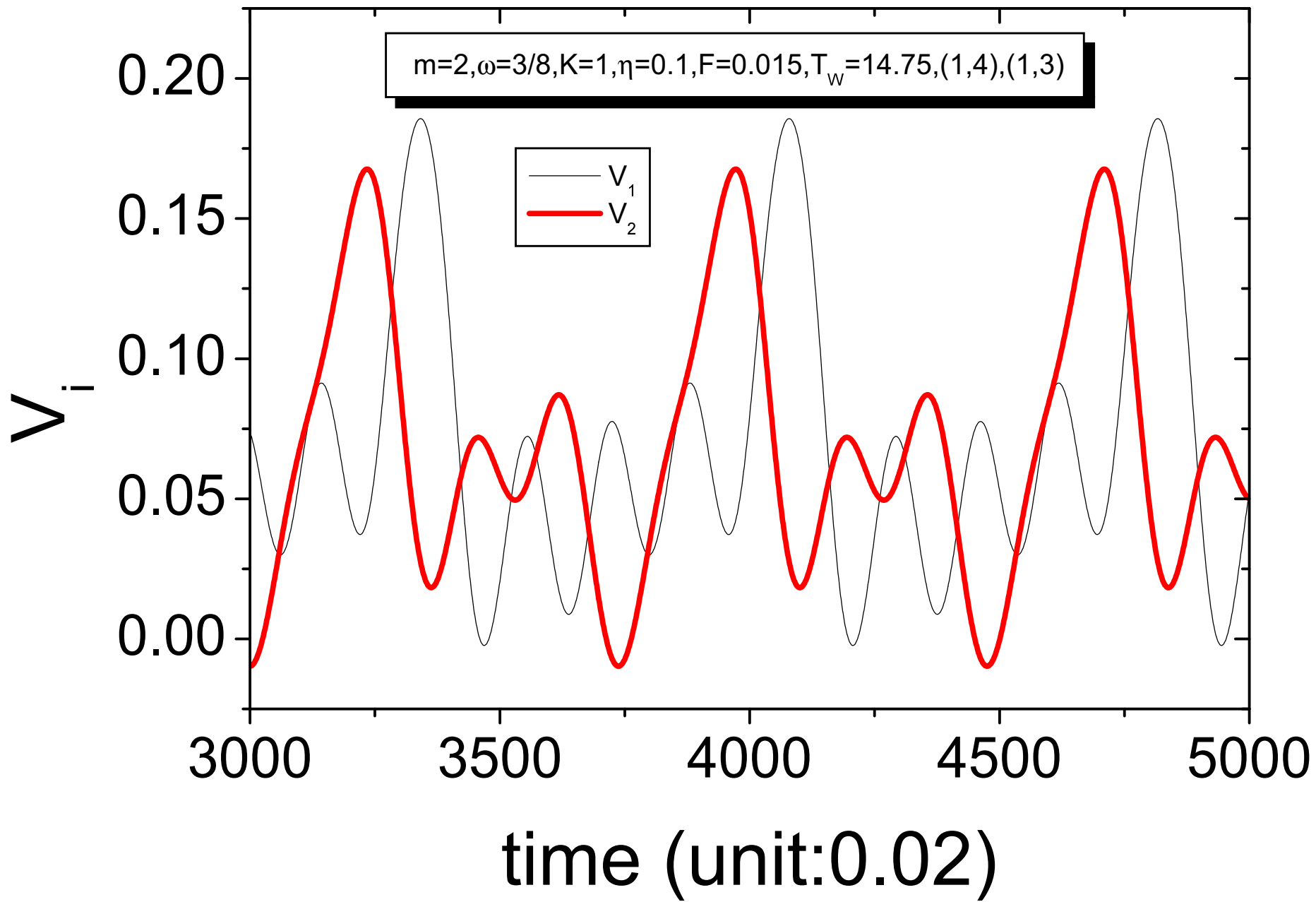
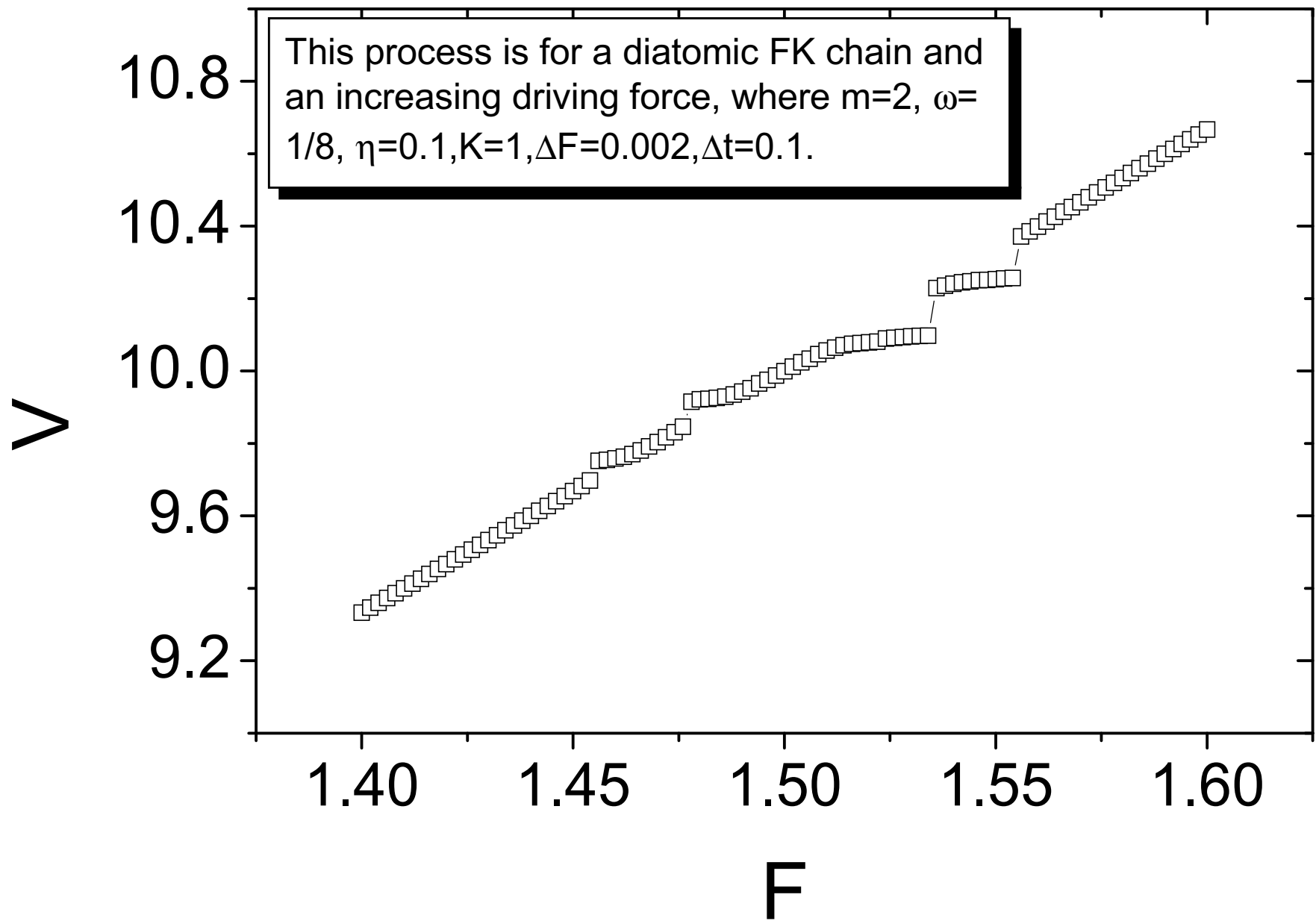
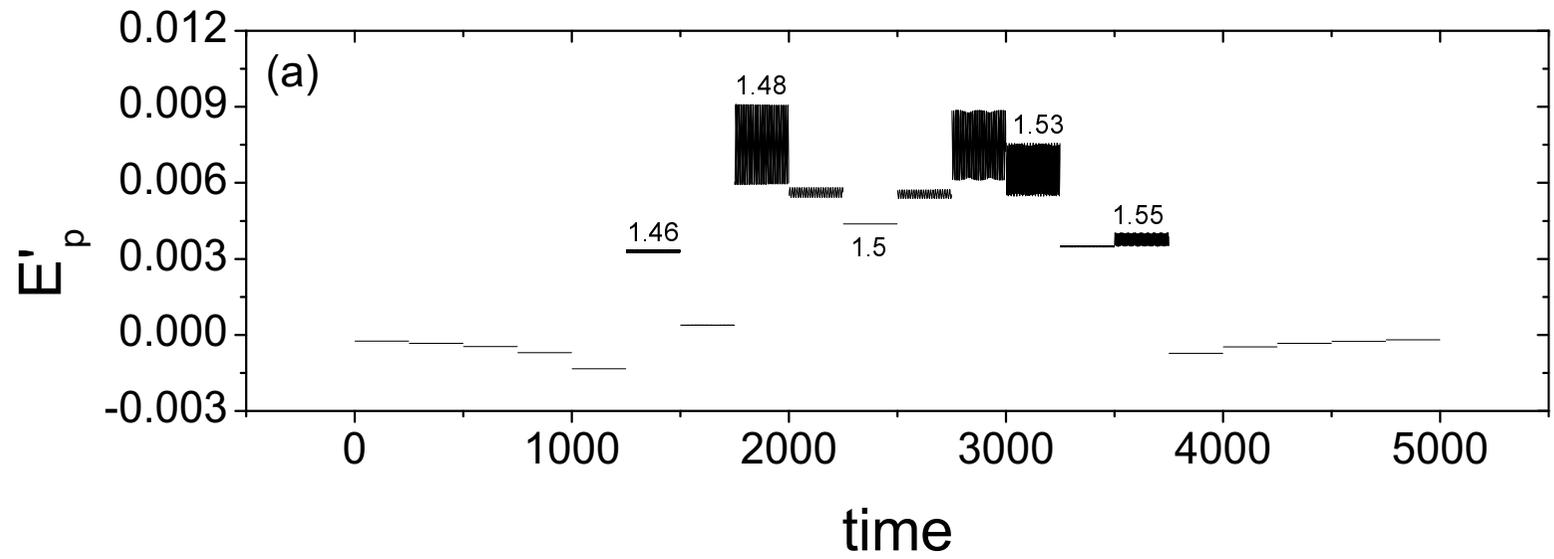
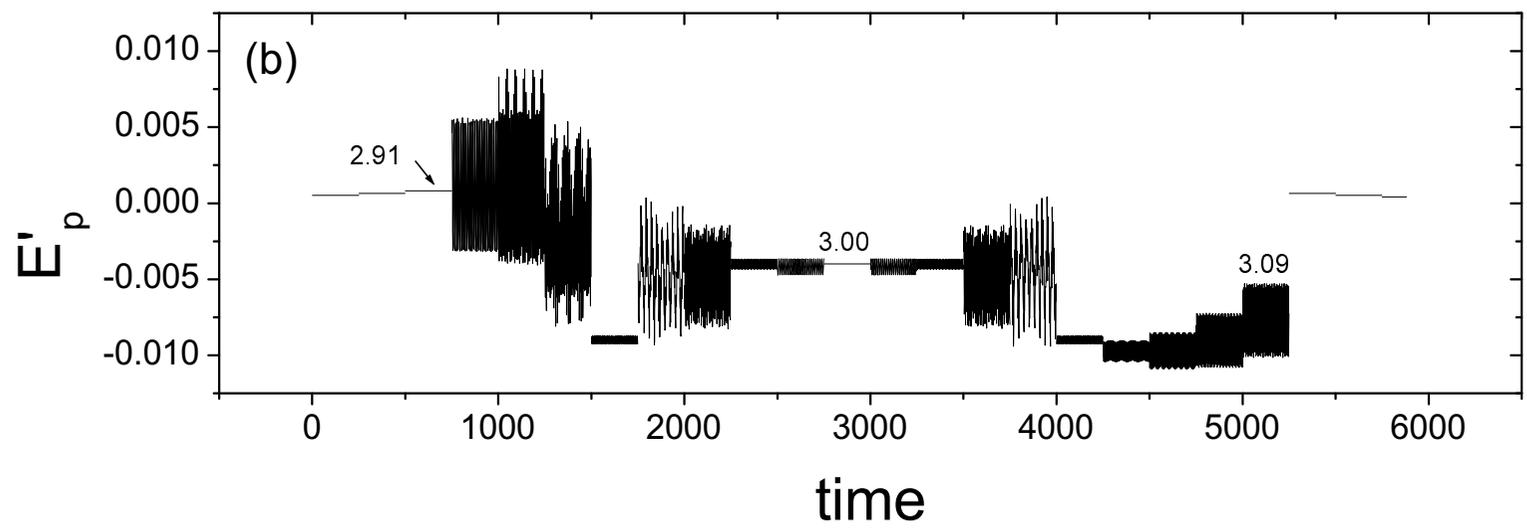
\end{document}